\documentclass[12pt]{iopart}

\usepackage{graphicx}
\begin{document}

\title{Particle production sources at LHC energies}

\author{Georg Wolschin}

\address{Institut f{\"ur} Theoretische 
Physik
der Universit{\"a}t Heidelberg, Philosophenweg 16, D-69120 Heidelberg, Germany, EU}
\ead{wolschin@uni-hd.de}
\begin{abstract}
Particle production sources at RHIC and LHC energies are investigated in pseudorapidity space.
A nonequilibrium-statistical relativistic diffusion model (RDM) with three sources is applied
to the analysis of charged-hadron distributions in AuAu collisions at RHIC energies, in PbPb collisions at the current LHC energy of 2.76 TeV, in $p$Pb at 5.02 TeV, and in $pp$. 
The size of the midrapidity source relative to the fragmentation sources in heavy-ion collisions
 is investigated as function of incident energy. At LHC energies, the midrapidity value is mostly determined by particle production from gluon--gluon collisions. 

\end{abstract}


\section{Introduction}

Charged-hadron production in relativistic heavy-ion collisions has been investigated in great detail at the Relativistic Heavy-Ion Collider RHIC in AuAu collisions, and more recently at the Large Hadron Collider LHC in PbPb collisions. In particular, high-precision pseudorapidity distributions $dN_{ch}/d\eta$ of produced charged particles including their centrality dependence are now available in an energy range from $\sqrt{s_{NN}} = 0.019$ to 2.76 TeV \cite{alv11,sa12}. At RHIC energies these data include the fragmentation regions up to the values of the beam rapidities, whereas at the current LHC energy of 2.76 TeV corresponding to a beam rapidity of 7.99 preliminary but very precise ALICE data are available up to $\eta\simeq 5$ \cite{sa12}.

Theoretical descriptions of the underlying partonic processes often focus on gluon-gluon production, such as in many approaches based on the color glass condensate (see \cite{alb07} as an example). Based on this mechanism particle and antiparticle distributions would, however, be identical -- which is not the case experimentally, as found for example in $\pi^+$ and $\pi^-$ distribution functions \cite{bea01}.

The relevance of the fragmentation sources from quark-gluon interactions has been investigated in a recent QCD-based study of net-baryon distributions (baryons minus antibaryons). There the gluon-gluon source that is peaked at midrapidity cancels out such that only the fragmentation sources remain \cite{mtw09,mtwc09}, giving rise to two fragmentation peaks that are clearly seen in the data at high SPS and RHIC energies, and in the theoretical predictions at LHC energies. At low SPS energies the fragmentation peaks overlap in rapidity space and hence, are not directly visible in the data, but can still be extracted quite reliably \cite{mtw11}.

For produced particles (rather than net baryons), the effect of the fragmentation sources is less obvious, but clearly has to be considered. In this note I propose to investigate the relative importance of gluon-gluon vs. fragmentation sources as a function of c.m. energy in collisions of heavy systems (AuAu, PbPb) using a phenomenological nonequilibrium-statistical model. This relativistic diffusion model (RDM) \cite{wol99} has proven to be useful in the analysis of data and in predictions for asymmetric \cite{wobi06} and symmetric \cite{rgw12} systems, and -- in spite of the small transverse size of the system -- also in $pp$ collisions \cite{gwo11}. Its three sources correspond to the gluon-gluon and fragmentation sources of the available microscopic theories. In direct comparisons with the available data the RDM can be used to infer the relative sizes of these underlying components as functions of the incident energy. 

In charged-hadron production at SPS and low RHIC energies up to $\sqrt{s_{NN}} \simeq 20$ GeV, the gluon-gluon source centered at midrapidity is expected -- and has turned out -- to be unimportant \cite{kw07}, and the measured pseudorapidity distributions are well reproduced from the fragmentation sources only. At these relatively low energies, the fragmentation sources are peaked close to midrapidity and hence, are influenced considerably by the Jacobian transformation from rapidity to pseudorapidity space. At higher energies, the fragmentation peaks move apart, and the central gluon-gluon source emerges. Then the Jacobian increasingly affects only the central source. Also, its overall effect becomes smaller with rising energy since it depends on $(\langle m \rangle/p_T)^2$. Still, a precise determination of the Jacobian is essential for the modeling of pseudorapidity distributions at LHC energies. The pronounced midrapidity dip that is seen in the recent ALICE PbPb charged-hadron data is due to the interplay of fragmentation and central sources, plus the effect of the Jacobian on the central source.

A brief outline of the method used to determine the relative size and extent of the sources in $\eta-$space is given in the next section.
Results for heavy systems at RHIC and LHC energies are presented in section 3. The energy dependence of central and fragmentation sources is discussed in section 4. A brief outlook on $p$Pb at 5.02 TeV and $pp$ at 0.9 -- 14 TeV is given in section 5, and the conclusions are drawn in section 6.

\begin{figure}
\centering
\includegraphics[width=11cm]{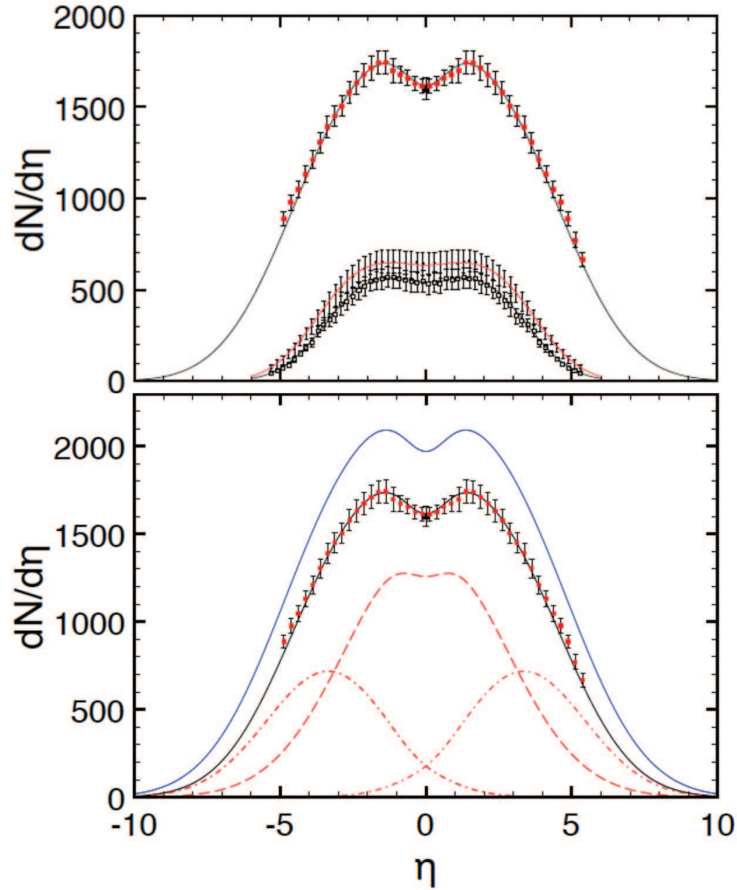}
\caption{
\label{fig1}
(Color online) The RDM  pseudorapidity distribution function for charged hadrons in 0--5\% central PbPb collisions
at LHC energies of 2.76 TeV is shown in the upper panel of the figure, with the
RDM parameters adjusted in a $\chi^2-$ minimization to the ALICE data (central value from \cite{aamo10},
distribution from \cite{sa12}). 
The fit takes the limiting fragmentation scaling hypothesis into account \cite{rgw12}. The corresponding 
RDM parameters are given in table~\ref{tab1}. 
In the lower part of the upper frame,
calculated pseudorapidity distributions of 
produced charged particles from
AuAu collisions (bottom) at $\sqrt{s_{NN}}$ =  0.13 and 0.2 TeV for
0--6\% central collisions with PHOBOS data \cite{alv11}
 are shown for comparison, see \cite{gw11}. In the bottom frame,  the underlying theoretical distributions
are shown for 2.76 TeV PbPb. Only the shape of the midrapidity source is modified by the Jacobian.
At LHC energies, the midrapidity value is mostly determined by particle production from gluon--gluon collisions. The upper curve is the RDM-prediction for 5.52 TeV.}
\end{figure}

\section{Three sources model}

In  the three-sources version of the relativistic diffusion model, rapidity distributions of produced particles
are calculated from an incoherent superposition of the fragmentation sources $R_{1,2}(y,t=\tau_{int})$ with charged-particle content $N_{ch}^{1}$ (projectile-like), $N_{ch}^{2}$ (target-like) and
the midrapidity gluon-gluon source $R_{gg}(y,t = \tau_{int})$ with charged-particle content $N_{ch}^{gg}$  as

\begin{equation}
\frac{dN_{ch}(y,t=\tau_{int})}{dy}=N_{ch}^{1}R_{1}(y,\tau_{int})
 +N_{ch}^{2}R_{2}(y,\tau_{int})
+N_{ch}^{gg}R_{gg}(y,\tau_{int})
\label{normloc1}
\end{equation}
with the rapidity $y = 0.5\cdot \ln((E+p)/(E-p))$, and the interaction time $\tau_{int}$ (total integration time of the underlying partial differential equation). 

In the linear version of the RDM \cite{wol99}, the macroscopic distribution functions are solutions of the Fokker-Planck equation ($k = 1, 2, 3$)
\vspace{.2cm}
\begin{equation}
\frac{\partial}{\partial t} R_{k}(y,t) =
-\frac{1}{\tau_{y}}\frac{\partial}
{\partial y}\Bigl[(y_{eq}-y)\cdot R_{k}(y,t)\Bigr]
+D_{y}^{k} \frac{\partial^2}{\partial y^2}
R_{k}(y,t).
\label{fpe}
\end{equation}
\vspace{.4cm}
The use of the Lorentz-invariant variable rapidity in the nonequilibrium-statistical Fokker-Planck framework has proven to be a useful approach in calculations and predictions of macroscopic distribution functions for produced particles.
Integrating the equation with the initial conditions
$R_{1,2}(y,t=0)=\delta(y\pm y_{max})$, the absolute value of the beam rapidities 
$y_{max}$, and $R_{3=gg}(y,t=0)=\delta(y-y_{eq})$  
yields the exact solution. The mean values 
are derived analytically from the moments 
equations as
\begin{equation}
<y_{1,2}(t)>=y_{eq}[1-\exp(-t/\tau_{y})] \mp y_{max}\exp{(-t/\tau_{y})}
\label{mean}
\end{equation}
for the sources (1) and (2) with the absolute value of the beam rapidity $y_{max}$ and the rapidity relaxation time $\tau_y$.

The local equilibrium value $y_{eq}$ is equal to zero only for
symmetric systems. For asymmetric systems such as $p$Pb, the midrapidity source is moving
\cite{wob06}, and the superposition of the sources is even more sensitive to
the values of the model parameters than in the symmetric case. From energy-momentum
conservation the centrality-dependent equilibrium value is obtained as \cite{bha53,nag84}
\begin{equation}
y_{eq}(b)=-0.5\cdot\ln{\frac{\langle m_1^{T}(b)\rangle\exp(y_{max})+\langle m_2^{T}(b)\rangle\exp(-y_{max})}
{\langle m_2^{T}(b)\rangle\exp(y_{max})+\langle m_1^{T}(b)\rangle\exp(-y_{max})}}
\label{eq}
\end{equation}
with the beam rapidities $y_{beam}=\mp y_{max}$, the average transverse masses $\langle m_{1,2}^T(b)\rangle =
\sqrt{m^2_{1,2}(b)+\langle p_T\rangle^2}$, and participant masses $m_{1,2}(b)$ of the $p$- and Pb-like participants in $p$Pb collisions that depend on the impact parameter $b$. The minus sign refers to cases where $m_2^T >m_1^T$ such as in the ALICE $p$Pb experiment of 2012 where the Pb beam defined the positive rapidity. The sign of the equilibrium value changes when the beams are interchanged, as is planned in the 2013 $p$Pb experiments at the LHC.

For sufficiently large beam rapidities $y_{max}$ such at LHC energies, the equilibrium value can be approximated as
\begin{equation}
y_{eq}(b)\simeq 0.5\cdot\ln{\frac{\langle m_2^{T}(b)\rangle}{\langle m_1^{T}(b)\rangle}}
\hspace{.2cm} .
\label{eq1}
\end{equation}

The corresponding numbers of participants can be obtained from the geometrical overlap, or from Glauber calculations. For $y_{beam}\simeq 8.6$ as in 5.02 TeV $p$Pb an estimate in $0-5\%$ central collisions is $y_{eq} \simeq 0.6$, and smaller values for more peripheral collisions. The time evolution in the RDM causes a drift of the distribution functions $R_{1,2}$ towards $y_{eq}$. 

Whether the mean values of $R_1$ and $R_2$ actually attain $y_{eq}$ depends on the centrality-dependent interaction time $\tau_{int}$ (the time the system interacts strongly, corresponding to the integration time of (\ref{fpe})), and its ratio to the rapidity relaxation time $\tau_y$. Typical interaction times at LHC energies from dynamical models in central
PbPb collisions are 6-8 fm/$c$, which is too short for the fragmentation sources to reach equilibrium, such that their mean values $<y_{1,2}>$ remain between beam and equilibrium values.  

This does not apply, however, to $R_{gg}$ which already emerges near equilibrium at the parton formation time -- here, at $t=0$ because of the (idealized) $\delta-$function initial condition -- and spreads in time due to strong diffusive interactions with other particles, without any shift in the mean value for a given centrality class.
The variances are
\begin{equation}
\sigma_{k}^{2}(t)=D_{y}^{k}\tau_{y}[1-\exp(-2t/\tau_{y})],
\label{var}
\end{equation}
so they reach equilibrium faster than the mean values. Here the diffusion coefficients in rapidity space are $D_y^k$, and presently I assume equal values for the three sources.
The corresponding FWHM-values $\Gamma_{1,2,gg}$ 
are listed in table~\ref{tab1} for AuAu and PbPb together with the mean values. Both are determined from $\chi^2-$ minimizations with respect to the very precise PHOBOS and preliminary ALICE data \cite{alv11,sa12}.

Since the theoretical model is formulated in rapidity space, one has to transform the calculated distribution functions to pseudorapidity space, $\eta=-$ln[tan($\theta / 2)]$, in order to be able to compare with the available data, and perform $\chi^2-$minimizations. The well-known Jacobian transformation\\ 
\begin{equation}
\frac{dN}{d\eta}=\frac{dN}{dy}\frac{dy}{d\eta}=
J(\eta,  m / p_{T})\frac{dN}{dy}, 
\label{deta}
\end{equation}
\begin{equation}
{J(\eta,m /p_{T})=\cosh({\eta})\cdot }
[1+(m/ p_{T})^{2}
+\sinh^{2}(\eta)]^{-1/2}
\label{jac}
\end{equation}
depends on the squared ratio of the mass and the transverse momentum of the produced particles.
Hence, its effect increases with the mass of the particles, and it is most pronounced at small transverse momenta. For reliable results one has to consider the full $p_T-$distribution, however: It is not sufficient to consider only the mean transverse momentum $\langle p_T \rangle$. In \cite{rgw12} we have discussed in some detail how this can be done for known $p_T-$distributions of identified $\pi^-, K^-$, and antiprotons. We use the pion mass $m_{\pi}$, and then calculate an effective mean transverse momentum $<p_T^{eff}>$ such that the experimentally determined Jacobian $J_{y=0}$ of the total charged-hadron distribution at rapidity  zero is exactly reproduced. This yields
for a given centrality class \cite{rgw12}
\begin{equation}
\langle p_T^{eff}\rangle=m_{\pi}J_{y=0}/\sqrt{1-J_{y=0}^2}\hspace{.2cm} .
\end{equation}
These effective transverse momenta are smaller than the mean transverse momenta determined from the $p_T-$distributions, and the corresponding effect of the Jacobian is therefore larger than that estimated
with $\langle p_T \rangle$ taken from the transverse momentum distributions for each particle species. At high RHIC and LHC energies the effect of the Jacobian transformation remains, however, essentially confined to the midrapidity source.

The Jacobians can now be calculated for each centrality class, pseudorapidity distributions of produced charged hadrons are obtained in the three-sources model from (\ref{normloc1}), the parameters are optimized with respect to the available data, and conclusions regarding the relative sizes of the sources become possible.

However, LHC data are still missing in the fragmentation region. We have therefore proposed in \cite{rgw12} to use the well-known limiting fragmentation scaling hypothesis \cite{ben69} as an additional constraint: At sufficiently high energy, particle production in the fragmentation region becomes almost independent of the collision energy. Hence we use 0.2 TeV AuAu results at RHIC -- where data in the fragmentation region are available -- to supplement the LHC 2.76 TeV PbPb data in analogous centrality classes at large values of pseudorapidity, shifting the latter by $\Delta y= y_{beam}^{LHC}- y_{beam}^{RHIC}=7.99-5.36=2.63$. The resulting RDM-parameters have physically reasonable dependencies on the c.m. energy and centrality, and their extrapolations -- in particular, to higher energies such as the LHC design energy of 5.52 TeV PbPb -- can readily be used for predictions. 
\begin{figure}
\begin{center}
\includegraphics[width=11cm]{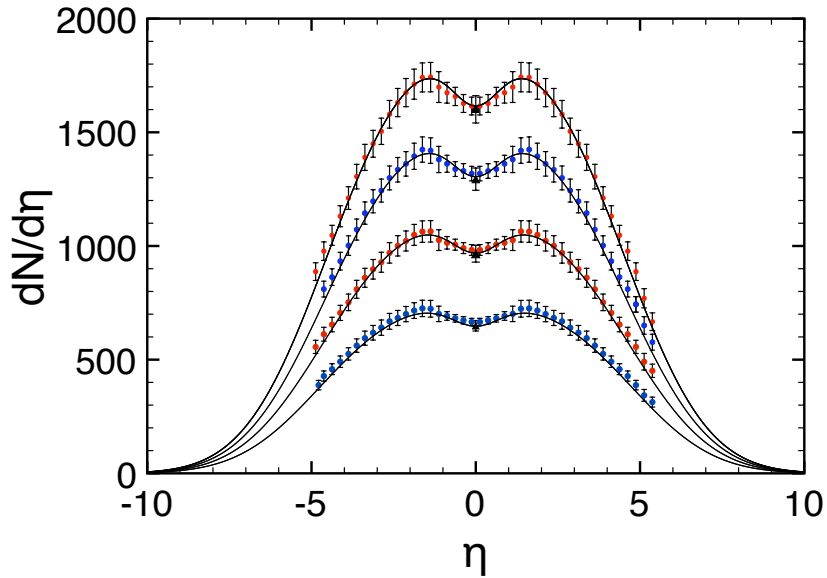}
\caption{\label{fig5}(Color online) Pseudorapidity distributions for produced charged hadrons in 2.76 TeV PbPb collisions as functions of centrality, from bottom to top: 20--30\%, 10--20\%, 5--10\%, 0--5\%. Calculated RDM distributions (solid curves) have been optimized in $\chi^2$-fits with respect to the preliminary ALICE data from \cite{sa12}, and using the limiting fragmentation scaling hypothesis in the region of large rapidities where no data are available.
The centrality-dependent parameter values are as in \cite{rgw12}. 
}
\label{fig2}
\end{center}
\end{figure}
In this work  the main emphasis is on the determination of the relative source sizes as function of the incident energy. For a reliable identification of the percentage of particles in the three sources the accurate modeling of the fragmentation region is essential, and limiting fragmentation scaling is a useful tool for PbPb collisions at LHC energies since accurate data at RHIC energies are available for AuAu at large values of pseudorapidity. 

For $p$Pb collisions, however, no low-energy data in the fragmentation regions exist, and limiting fragmentation scaling can not be used at LHC energies. Also, the presently available ALICE data at 5.02 TeV are limited to the range $-2<\eta<2$ \cite{ab12}.
It is therefore advisable to wait for the forthcoming 2013 LHC $p$Pb data in the pseudorapidity region $2<\eta<5$ before more definite conclusions regarding the relative source sizes can be drawn. 

For $pp$ at LHC energies there is presently also a lack of data in the fragmentation region. The $pp$ data at 0.9, 2.36 and 7 TeV \cite{kh10,kha10}  have a pseudorapidity coverage $ |\eta| < 2.25$.
There exist TOTEM inelastic results at 7 TeV for pseudorapidities 5.4--6.4 \cite{tot12}, but these are still far from the beam rapidity, and their normalization is not consistent with the  non-single diffraction (NSD) CMS results at midrapidity. As a consequence, the determination of the relative source sizes as function of the incident energy is restricted in this work to heavy-ion collisions, although I shall also present a $p$Pb analysis at 5.02 TeV, and $pp$ RDM results from the earlier work \cite{gw11} in comparison with the recent TOTEM data.

\section{Results}

The result of the three-sources RDM calculation for the pseudorapidity distribution of produced charged hadrons 2.76 TeV PbPb is shown in figure \ref{fig1} together with recent preliminary ALICE data \cite{sa12} for $0-5\%$ centrality in a $\chi^2$ optimization. Parameters are given in table \ref{tab1}. The published midrapidity ALICE data point \cite{aamo10} is slightly below, but within the error bars compatible with the more recent data. 

The three-sources fit uses the limiting fragmentation scaling assumption based on the 0.2 TeV AuAu central PHOBOS data \cite{alv11} from RHIC, which are also shown in figure \ref{fig1} together with the 0.13 TeV AuAu data, and the corresponding RDM results \cite{gw11}, with parameters given in table \ref{tab1}. Results for central AuAu collisions at RHIC energies of 19.4 GeV and 62.4 GeV that are included in table \ref{tab1} are taken from earlier work \cite{rgw12,gw11}.
A prediction for the LHC design energy of 5.52 Te PbPb is also shown in figure \ref{fig1}.

The relative size of the three sources in central 2.76 TeV PbPb is shown in the lower frame of figure \ref{fig1}. At this LHC energy, the midrapidity source already contains the largest fraction of produced charged hadrons. Its shape is significantly deformed by the Jacobian transformation from rapidity to pseudorapidity space, whereas the fragmentation sources are not much influenced by the transformation.

In the full distribution that arises from the incoherent superposition of the three sources, it is evident that the midrapidity dip is more pronounced at LHC energies as compared to RHIC energies, although the effect of the Jacobian tends to be smaller at the higher incident energy. This clearly indicates that there has to be a physical origin of the midrapidity dip in addition to the effect of the Jacobian. 

The hypothesis promoted in this work is that the interplay of the three sources provides the observed effect. In 2.76 TeV PbPb collisions, the fragmentation sources are peaked at large values ($<y_{1,2}>=3.34$) of rapidity -- whereas at 0.2 TeV RHIC energy, the center is at $<y_{1,2}>=2.4$. Consequently,  the midrapidity yield at LHC energies is essentially due to the central source, with only a small contribution from the fragmentation sources. Although the relative particle content in the central source is larger at LHC energies than at RHIC, this produces the observed midrapidity dip, together with the effect of the Jacobian on the central source.

The centrality dependence of charged hadron production at LHC energies as displayed in figure \ref{fig2} in comparison with preliminary ALICE PbPb data for 0-5\%, 5-10\%, 10-20\% and 20-30\% shows that the total number of produced charged hadrons rises with increasing centrality. As displayed in figure 2 of \cite{rgw12} that was based on earlier preliminary ALICE data, the rise is almost linear with increasing number of participants. The percentage of particles in the midrapidity source depends weakly on centrality, falling from 56 \% in central collisions to 51 \% at 20-30 \% centrality.

\begin{figure}
\begin{center}
\includegraphics[width=12cm]{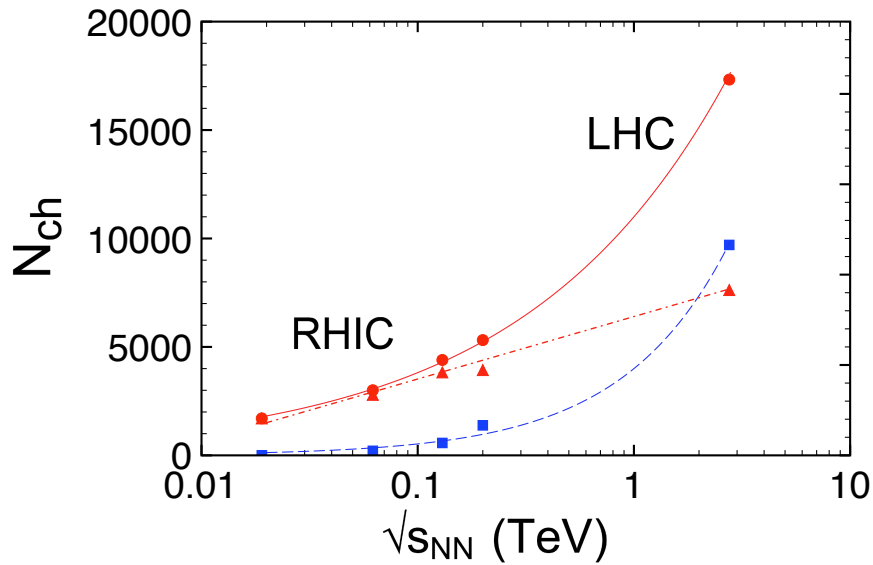}
\caption{\label{fig3}(Color online) Number of produced charged hadrons as function of the c.m. energy $\sqrt{s_{NN}}$ from  RDM-fits of the available data for central heavy-ion collisions at 0.019, 0.062, 0.13, 0.2 TeV (RHIC, AuAu), and 2.76 TeV (LHC, PbPb). Circles are the total numbers, following a power law $\propto s_{NN}^{0.23}$. Squares are hadrons produced from the midrapidity source, with a power law $\propto s_{NN}^{0.44}$, and triangles are particles from the fragmentation sources $\propto\log(s_{NN}/s_0)$. 
The gluon-gluon source (dashed) becomes the main source of particle production between RHIC and LHC energies.
}
\end{center}
\end{figure}

\begin{table*}
\begin{center}
\caption{\label{tab1}Three-sources RDM-parameters $\tau_{int}/\tau_{y}$, $\Gamma_{1,2},  \Gamma_{gg}$, and $N_{gg}$. $N_{ch}^{1+2}$ is the total charged-particle number in the fragmentation sources, $N_{gg}$ the number of charged particles produced in the central source.
Results for $<y_{1,2}>$ are calculated from
 $y_{beam}$ and $\tau_{int}/\tau_{y}$. 
Values are shown
for 0--5\% PbPb at LHC energies of 2.76 and 5.52 TeV in the lower two lines,  with results at 2.76 TeV from a $\chi^2$-minimization with respect to the preliminary ALICE data \cite{sa12}, and using limited fragmentation as constraint. 
Corresponding parameters
for 0--6\% AuAu at RHIC energies are given for comparison in the upper four lines based on PHOBOS results \cite{alv11}.
Parameters at 5.52 TeV denoted by * are extrapolated.  
Experimental midrapidity values (last column) are from PHOBOS \cite{alv11} for $|\eta| < 1$, 0-6\% at RHIC energies and from ALICE \cite{aamo10}  for $|\eta| < 0.5$, 0-5\% at
2.76 TeV.} 
\vspace{.3cm}
\begin{tabular}{lllllllcr}
\hline\\
$\sqrt{s_{NN}} $&$y_{beam}$& $\tau_{int}/\tau_{y}$&$<y_{1,2}>$&$\Gamma_{1,2}$&$\Gamma_{gg}$&$N_{ch}^{1+2}$&$N_{gg}$&$\frac{dN}{d\eta}|_{\eta \simeq 0}$\\
   (TeV)\\
\hline\\
 0.019&$\mp 3.04$&0.97&$\mp 1.16$& 2.83&0&1704&-&314$\pm 23$\cite{alv11}\\
     0.062&$\mp 4.20$&0.89&$\mp 1.72$& 3.24&2.05&2793&210&463$\pm 34$\cite{alv11}\\  
  0.13&$\mp 4.93$&0.89&$\mp 2.02$& 3.43&2.46&3826&572&579$\pm 23$\cite{alv11}\\
 0.20&$\mp 5.36$&0.82&$\mp 2.40$& 3.48&3.28&3933&1382&655$\pm 49$ \cite{alv11}\\
 2.76&$\mp 7.99$&0.87& $\mp 3.34$&4.99&6.24&7624&9703&1601$\pm 60$ \cite{aamo10}\\
  5.52&$\mp 8.68$&0.85*& $\mp 3.70$&5.16*&7.21*&8889*&13903*&1940*\\\\
\hline 
\end{tabular}
\end{center}
\end{table*}

\section{Energy dependence of the hadron production sources}

There are now sufficiently precise data on charged-hadron production at RHIC \cite{alv11} and LHC \cite{sa12}  energies available in order to investigate the relative size of the three particle production sources as function of energy in heavy-ion collisions (AuAu at RHIC, PbPb at LHC). Based on $\chi^2$ optimizations of the analytical three-sources RDM solutions with respect to these data I have displayed the energy dependence of the sources in figure \ref{fig3}, with parameters as shown in table \ref{tab1}.

According to these results, the total charged-hadron production (circles) follows a power law $\propto s_{NN}^{0.23}$. The hadrons produced from the central source (squares) have an even stronger dependence on initial energy according to $\propto s_{NN}^{0.44}$, whereas particles produced in the fragmentation sources have a weaker dependence $\propto\log(s_{NN}/s_0)$.

The strong rise of the particle production yield from the central (gluon-gluon induced) source is evidently due to the increasing gluon content of the system at high relativistic energies. In particular, the total particle production rate from the central source becomes larger than that from the two fragmentation sources at an incident energy between the highest RHIC energy (0.2 TeV), and the LHC regime. In view of the lack of data in this intermediate regime, the precise crossing point is, however, difficult to determine. 
\begin{figure}
\centering
\includegraphics[width=11cm]{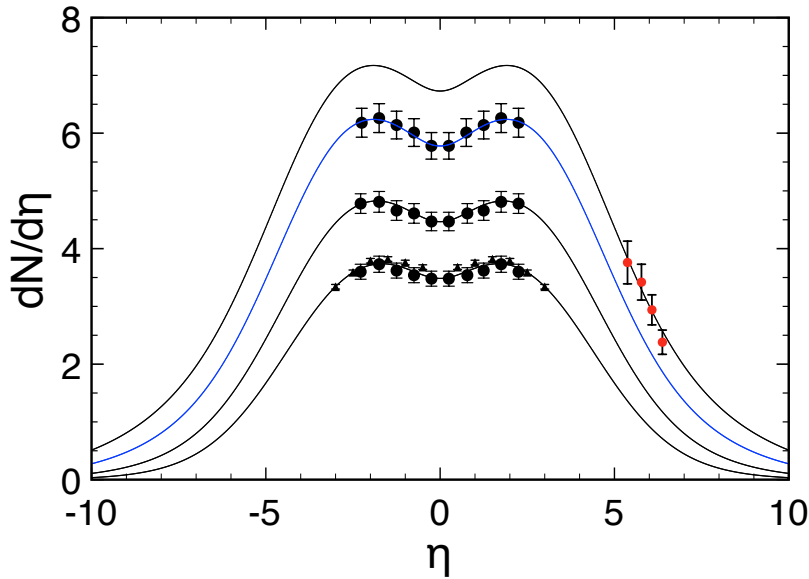}
\caption{Pseudorapidity distributions of produced 
charged hadrons in $pp$ collisions (NSD) at LHC c.m. energies of 0.9, 2.36, 7 and 14 TeV (bottom to top) as calculated in the three-sources approach \cite{gwo11} and fitted to CMS NSD data \cite{kh10,kha10}.  At 0.9 TeV UA5 NSD data are also shown \cite{an89}, triangles. Data points in the region $\eta = 5.375$ to 6.375 are results from the TOTEM collaboration \cite{tot12} at 7 TeV (95\% inelastic; the prediction was normalized to NSD).} 
\label{fig4}
\end{figure}

\section{Proton-induced collisions}


I had previously applied the relativistic diffusion model \cite{wol99} also to $pp$ collisions at RHIC and LHC energies \cite{gwo11}. Here transport phenomena are not expected to be fully developed because of the small transverse size of the system, but still, the RDM yields reasonable results for pseudorapidity distributions of produced charged hadrons in the $\eta-$range where data are available, and provides predictions for large values of $\eta$. 

The TOTEM collaboration has reported first experimental $pp$ results at 7 TeV and large $|\eta|$ \cite{tot12}. Measured values for $dN/d\eta$ of charged hadrons range from 3.84$\pm$0.01(stat)$\pm$0.37(syst) at $|\eta| = 5.375$ to 2.38$\pm$ 0.01(stat)$\pm$0.21(syst) at $|\eta| = 6.375$. The data account for about 95\% of the total inelastic cross section. A corresponding RDM prediction taken from \cite{gwo11} is compared with these new data in figure \ref{fig4}. It has the correct slope, whereas MC predictions give a different slope \cite{tot12}. The absolute magnitude of the RDM result is, however, slightly too low since it was normalized to 7 TeV CMS NSD $pp$ data at midrapidity, which are below inelastic results. 

In order to draw definite conclusions regarding the particle content of the three sources in $pp$ collisions as function of $\sqrt s$ one would need data at various incident energies over a larger range in pseudorapidity than what is presently available. The same is true for charged-hadron production in asymmetric proton-induced collisions. Here ALICE data from the 2012 testrun are available for pseudorapidity distributions of produced charged hadrons in $p$Pb collisions at $\sqrt{s_{pN}}=5.02$ TeV (corresponding to a proton beam momentum of 4 TeV/$c$)  in the range $-2<\eta<2$ \cite{ab12}. 

The midrapidity yield is almost two orders of magnitude smaller than in 0-5\% central PbPb collisions at 2.76 TeV, and the shape of the distribution function is very sensitive to the details of the underlying partial distributions, see figure \ref{fig5}. The central source is seen to be significantly modified by the Jacobian, whereas the asymmetric fragmentation sources are almost gaussian-shaped. 

The interplay of the fragmentation sources with the moving gluon-gluon source that is centered at small positive $\eta$ values generates the characteristic shape of the charged-hadron distribution. The proton-like side of the distribution is seen to be considerably steeper than the Pb-like side. It is expected that the 2013 $p$Pb runs with interchanging beams will enable the LHC heavy-ion experiments to actually measure this predicted difference. A corresponding effect had been observed in $d$Au at 0.2 TeV \cite{alv11,wobi06}. There it was also shown that the difference in the slopes becomes more pronounced with increasing centrality, and a similar behaviour is expected at LHC energies once centrality-dependent data become available.

For a reliable determination of the particle content in the three sources one needs data at larger values of pseudorapidity, which are not yet available. Moreover, for $p$Pb only a single incident energy is available, so that an investigation of the relative particle content in the three sources as function of $\sqrt{s_{pN}}$ similar to the heavy-ion case is presently not feasible.
\begin{figure}
\centering
\includegraphics[width=12cm]{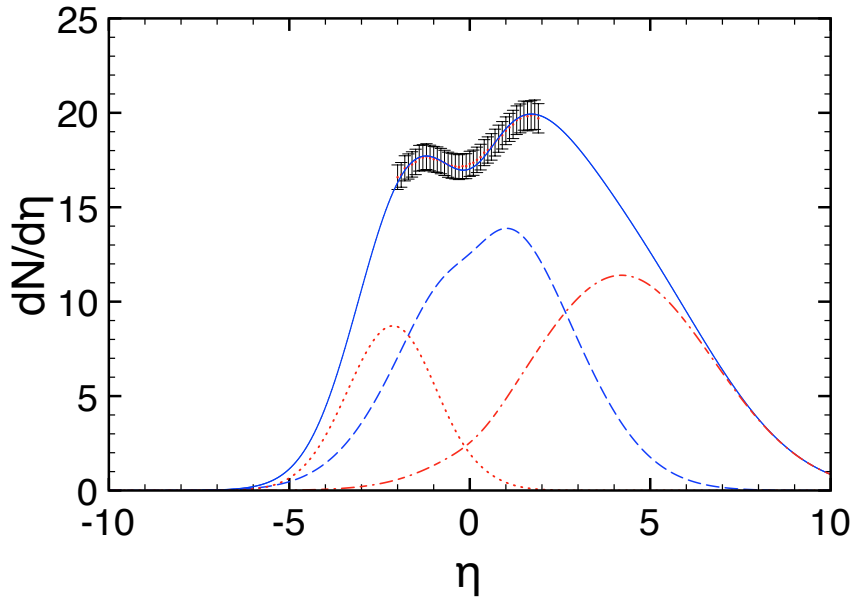}
\caption{
\label{fig5}
(Color online) The predicted RDM pseudorapidity distribution function for charged hadrons in minimum bias $p$Pb collisions
at LHC c.m. energy of 5.02 TeV shown here is adjusted in the mid-rapidity region to the preliminary ALICE data  \cite{ab12} (systematic error bars only).
The underlying distributions in the three--sources RDM are also shown, with the dashed curve
arising from gluon-gluon collisions, the dash-dotted curve from valence quark-gluon events in the Pb-like region, and the dotted curve correspondingly in the proton-like region. 
}
\end{figure}

\section{Conclusions}

The particle content of fragmentation and midrapidity (gluon-gluon) sources for charged-hadron production in heavy-ion collisions at high relativistic energies has been determined as function of c.m. energy in a phenomenological approach.
Due to the availability of preliminary high precision PbPb ALICE data from LHC \cite{sa12}, earlier AuAu PHOBOS data \cite{alv11}, and with the limiting fragmentation scaling hypothesis in the region of large 
pseudorapidities where LHC data are still missing, a determination of the particle content in a three-sources non-equilibrium statistical model has become possible.  

Whereas the fragmentation sources are found to depend on energy $\propto\log(s_{NN}/s_0)$, the central source has a much stronger energy dependence $\propto s_{NN}^{0.44}$, and the total number of produced charged hadrons -- which arises from an incoherent superposition of the three sources -- behaves like  $\propto s_{NN}^{0.23}$. As a consequence, particle production from the gluon-gluon source becomes more important than that from the fragmentation sources in the energy range between the maximum RHIC energy of 0.2 TeV, and the current LHC energy of 2.76 TeV. 

The same approach has also been applied to charged-hadron production in $pp$ collisions at LHC energies. This yields predictions for the pseudorapidity distributions of produced charged hadrons in the region of large $\eta$ which are in reasonable agreement with recent inelastic data of the TOTEM collaboration \cite{tot12} for 7 TeV $pp$, although there is a normalization problem when comparing with the midrapidity NSD results. 
Due to the lack of large--$\eta$ data at other LHC energies, a precise determination of the particle content of the sources as function of c.m. energy is, however, presently not yet feasible in proton-proton collisions, although there exist very precise PHOBOS $pp$ data at RHIC energies of 0.2 and 0.41 TeV \cite{alv11} that extend into the fragmentation region.

In central $p$Pb collisions at the LHC c.m. energy of 5.02 TeV, ALICE data \cite{ab12} have been used to compare with the analytical solutions of the relativistic diffusion model (RDM). The shape of the pseudorapidity distribution is found to be very sensitive to the interplay of the $p-$ and Pb-like fragmentation sources, and the moving central source that is significantly modified by the Jacobian. The RDM calculation exhibits a steeper slope on the proton-like side, as compared to the Pb-like side. It is expected that the forthcoming LHC $p$Pb experiments confirm this behaviour.

\vspace{.4cm}
\bf{Acknowledgments}\\
\rm

I am grateful to the ALICE Collaboration for their preliminary as well as for their published results.
This work has been supported by the ExtreMe Matter Institute EMMI.


\section*{References}
\bibliographystyle{jphysg}

\bibliography{gw_prc_nt}

\end{document}